\documentclass[twocolumn,showpacs,preprintnumbers,amsmath,amssymb]{revtex4-1}
\usepackage{url}
\usepackage{graphicx}
\usepackage{color}
\usepackage{dcolumn}
\usepackage{bm}
\usepackage{verbatim}
\usepackage{subfigure}
\usepackage{rotating}
\usepackage{afterpage}
\usepackage[
unicode=true,%
colorlinks=true,%
linkcolor=blue,%
urlcolor=blue]{hyperref}

\newcommand{\pt}{p_{T}}

\begin{document}
\preprint{APS/123-QED}

\definecolor{red}{rgb}{1,0,0}
\title{Possible Evidence for Radial Flow of Heavy Mesons in d+Au Collisions}
\author{Anne M. Sickles}
\email{anne@bnl.gov}
\affiliation{
Department of Physics
Brookhaven National Laboratory
Upton, NY 11973
}

\date{\today}

\begin{abstract}
Recent measurements of particle correlations and the spectra of hadrons at both RHIC and
the LHC are suggestive of hydrodynamic behavior in very small collision systems (p+Pb, d+Au 
and possibly high multiplicity p+p collisions at the LHC).  
The measurements in p+Pb and d+Au collisions are both qualitatively and quantitatively similar to
what is seen in heavy ion collisions where low viscosity hot nuclear matter is formed.
While light quarks and gluons are thought to make up the bulk matter, one of the most 
surprising results in heavy ion collisions is that charm quarks also have a large $v_2$.
Measurements of the transverse momentum spectra of electrons from the decay of $D$ and $B$ mesons in d+Au   
collisions
show an enhancement in central collisions relative to p+p collisions.  We employ the blast-wave model
to determine if the flow of heavy quarks in d+Au and p+Pb collisions is able to explain the enhancement
observed in the data.  We find a reasonable description of the data with blast-wave parameters
extracted from fits to the light hadron spectra, suggesting hydrodynamics as a possible explanation.
\end{abstract}

\pacs{}
%\keywords{Suggested keywords}
\maketitle

\section{Introduction} \label{sec:intro} 
The aim of the heavy ion physics programs at RHIC and the LHC is to produce and study
the very hot dense strongly interacting matter produced in these collisions, the Quark Gluon Plasma (QGP).
There has been enormous success in describing the bulk properties of this matter with 
hydrodynamics (for a recent review see Ref.~\cite{Heinz:2013th}).  
Even more interesting, the ratio of sheer viscosity to entropy density, $\eta/s$
used in the hydrodynamic calculations is constrained by the 
data~\cite{Adare:2011tg,Gale:2012in,Luzum:2012wu}
to be very small and within a few times $1/4\pi$, the conjectured quantum lower
bound~\cite{Kovtun:2004de}.

One of the most interesting recent developments in heavy ion physics is the possibility
of collective behavior in very small systems.  This was first explored with the 
elliptic flow, $v_2$, of charged hadrons as measured by the ALICE and ATLAS 
collaborations~\cite{Abelev:2012ola,Aad:2012gla}.  Similar, but
slightly larger $v_2$ was found by the PHENIX collaboration in d+Au collisions
at RHIC~\cite{Adare:2013piz}.  This is in agreement with hydrodynamic calculations
which predicted a larger $v_2$ in d+Au collisions than in p+Pb collisions due to the
larger initial state eccentricity driven by the shape of the 
deuteron~\cite{Bozek:2011if,Bzdak:2013zma,Qin:2013bha}.  
While the hydrodynamic descriptions of the data are intriguing, other models such
as the Color Glass Condensate~\cite{Dusling:2013oia} 
have also been invoked to explain the observed 
correlations.  

Hydrodynamic behavior can also be inferred through the shape of the identified particle
transverse momentum ($\pt$) 
spectra~\cite{VanHove:1982vk}.  The ALICE and CMS collaborations have recently published
analyses of the spectra in p+Pb collisions~\cite{Chatrchyan:2013eya,Abelev:2013haa} 
that show an
increase in the $\langle \pt \rangle$ as a function
of the charged particle multiplicity in the event, behavior consistent with increasing radial 
flow with increasing event multiplicity~\cite{Bzdak:2013lva,Bozek:2013ska}.  

Another method to extract possible radial flow information from identified particle spectra is with the
blast-wave model~\cite{Kolb:2000fha,Schnedermann:1993ws}.  
This model assumes thermalization and expansion with a common velocity field.
Extractions of the parameters in p+Pb collisions show increasing flow
velocity, $\beta$, with increasing charged particle multiplicity~\cite{Abelev:2013haa}.
Interestingly, similar behavior has been observed by the STAR collaboration in d+Au 
collisions~\cite{Abelev:2008ab}.  The $\langle \beta \rangle$ observed in the most
central 20\% of d+Au collisions is consistent with the $\langle \beta \rangle$ observed in 50-60\% central 
Au+Au collisions~\cite{Abelev:2008ab}.
At that centrality the charged particle $v_2$ is large~\cite{Adams:2004bi,Adare:2010ux}
which is taken to be evidence of hydrodynamic behavior.
Blast-wave fits have also been
performed in p+p collisions.  In $\sqrt{s}=$~200~GeV p+p collisions the extracted
$\langle  \beta \rangle$ values are much smaller: 0.244$\pm$0.081~\cite{Abelev:2008ab}
and 0.28$\pm$0.02~\cite{Adare:2011vy}.  A calculation using a modified version of the blast-wave
model~\cite{Tang:2008ud}
finds $\langle \beta \rangle =$0.000$+$0.124~\cite{Jiang:2013gxa}. At higher
collision energies, above 900~GeV, $\langle \beta \rangle$ is found to increase within this model. 
Blast-wave parameters have been studied in multiplicity selected p+p collisions at $\sqrt{s}$~=~7~TeV
and similar $\langle \beta \rangle$ values are observed in the highest multiplicity p+p and
p+Pb collisions at $\sqrt{s_{NN}}$~=~5.02~TeV~\cite{Preghenella:2013vna}. 
A quantitative explanation of this similarity has not been put forward. Unfortunately, multiplicity
selected p+p collisions have not been studied at $\sqrt{s}$~=~200~GeV, so a similar comparison
at RHIC is not yet possible, but would be extremely informative.

Heavy quarks, charm and bottom, also appear to be affected by the presence of the
matter in heavy ion collisions.  At high $\pt$ in Au+Au
collisions there is observed to be substantial energy loss of heavy quark jets~\cite{Nguyen:2012yx} and
suppression of
heavy mesons~\cite{ALICE:2012ab,Tlusty:2012ix} and electrons from the decay of heavy 
hadrons~\cite{Adare:2006nq,Adare:2010de,Aggarwal:2010xp}.  
In p+p collisions most of these electrons are from heavy meson decay. 
Here, we neglect the small contributions to the electron yield from
the decays of $c$ and $b$ baryons.
At lower $\pt$ significant
$v_2$ of the electrons from heavy meson decay is observed in Au+Au 
collisions~\cite{Adare:2006nq,Adare:2010de}.
The STAR collaboration has observed
a low $\pt$ enhancement of $D$ mesons~\cite{Tlusty:2012ix} 
which has been described by calculations incorporating hydrodynamic 
behavior~\cite{He:2012df,Gossiaux:2012ea} and the data are well described by
a blast-wave calculation~\cite{Adare:2013wca,Batsouli:2002qf}.

In d+Au collisions, the PHENIX Collaboration has measured the yield of electrons
from the decays of heavy hadrons in d+Au collisions at $\sqrt{s_{NN}}$~=~200~GeV.
Relative to expectations from binary scaled p+p collisions the yield of these
electrons is enhanced by approximately 40\% in the 20\% most central d+Au collisions~\cite{Adare:2012qb}.
The origin of this effect is not understood, though it is consistent with 
a Cronin enhancement~\cite{Cronin:1974zm,Antreasyan:1978cw} which increases with the particle mass.
Inspired by the success of a hydrodynamic description of p+Pb and d+Au collisions,
in this work we investigate whether a blast-wave calculation constrained to the $\pi$,
$K$, $p,\bar{p}$ spectra in d+Au collisions at $\sqrt{s_{NN}}$=~200~GeV
 can explain the observed enhancement of heavy flavor decay electrons.  We also provide calculations
for central p+Pb collisions at 5.02~TeV.

\section{Method}

\begin{figure}
\includegraphics[width=\columnwidth]{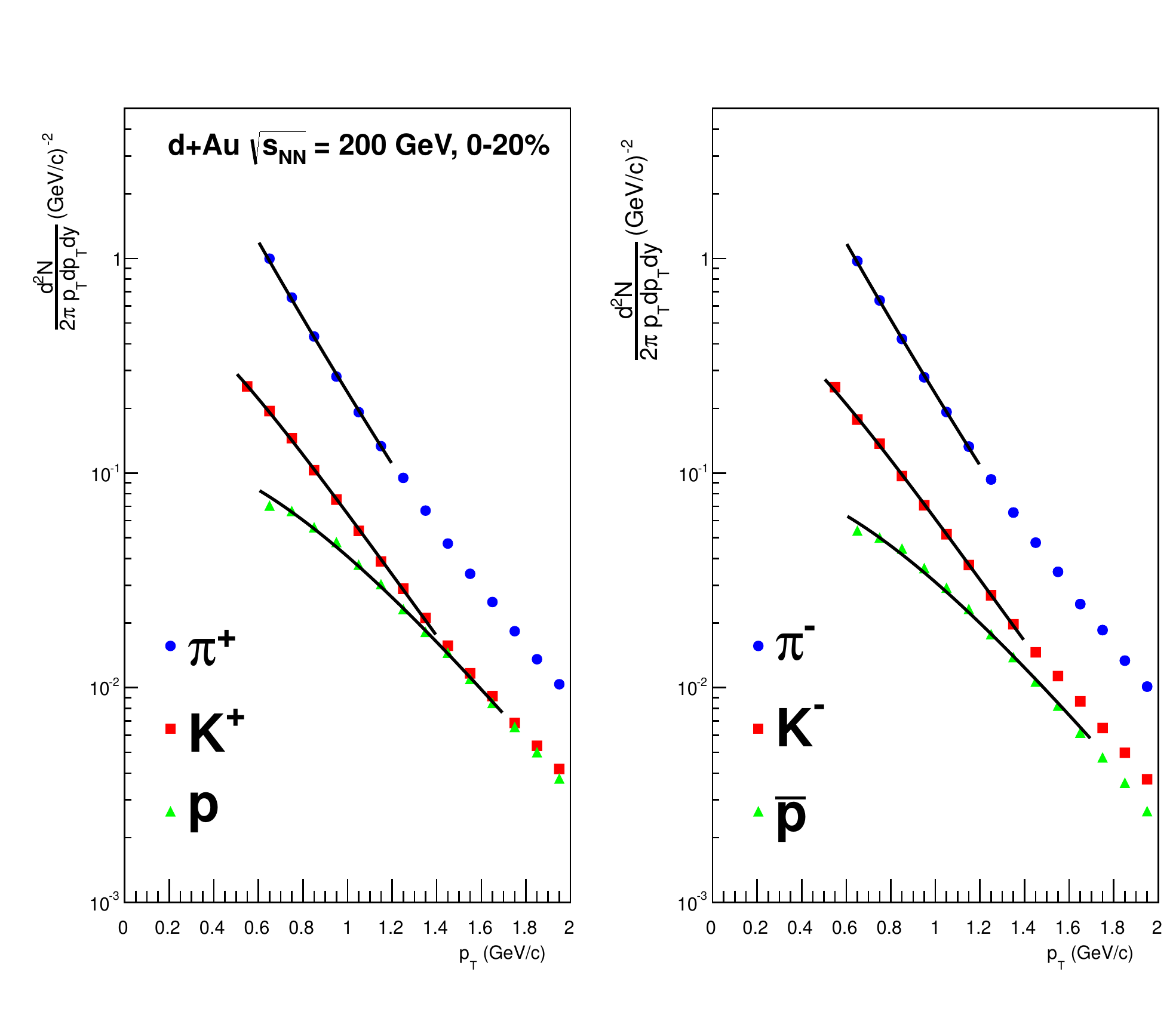}
\includegraphics[width=\columnwidth]{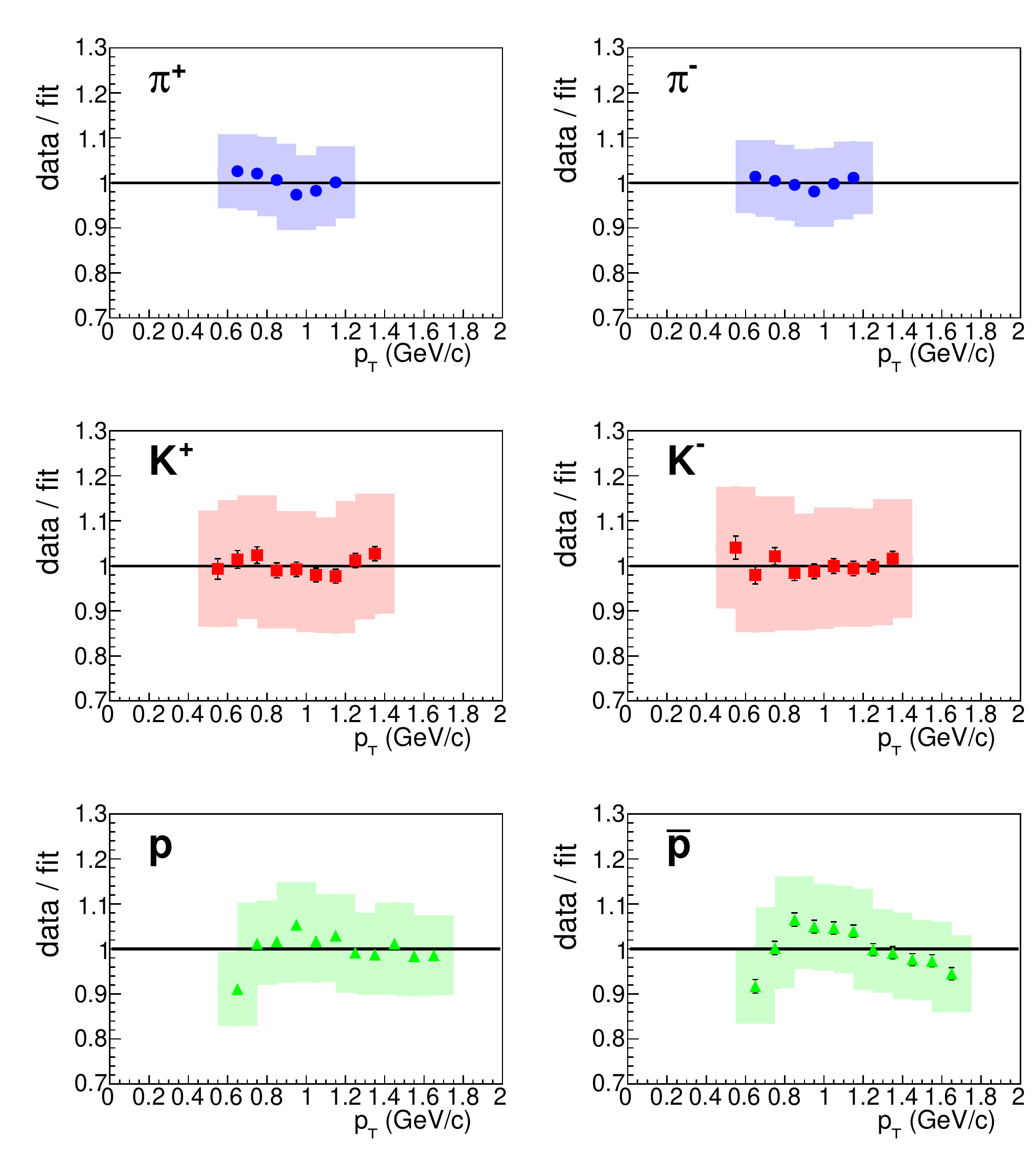}
\caption{(top panels) Charged hadron spectra for the 0-20\% most central d+Au 
collisions~\cite{Adare:2013esx}.  Overlaid with the data are the results
of a simultaneous blast-wave fit to the data. (bottom panels) Ratios 
of the experimental data to the blast-wave fits.  Statistical uncertainties on
the points are shown as error bars and systematic uncertainties are shown as boxes.}
\label{fig:spectra}
\end{figure}

The blast-wave model~\cite{Kolb:2000fha,Schnedermann:1993ws} 
describes $\pt$ spectra with the following functional form:
\begin{equation}
\frac{1}{\pt}\frac{dN}{d\pt} \propto
\int_{0}^{R} r \,dr \, m_{T} \, I_{0}\left(\frac{\pt \sinh \rho}{T_{fo}}\right) 
K_{1}\left(\frac{m_{T} \cosh \rho}{T_{fo}}\right)
\label{eq:bw}
\end{equation}
where:
\begin{equation}
\rho = \tanh^{-1} \left( \beta_{max} \left( r / R \right)^{n} \right)
\end{equation}
and $m_{T} = \sqrt{\pt^2 + m^2}$ where $m$ is the particle mass.
The model parameters are $\beta_{max}$, the maximum velocity at the surface,
and $T_{fo}$, the temperature at which the freeze out occurs.
We extract blast-wave parameters from $\pi^{\pm}$, $K^{\pm}$, $p$, $\bar{p}$ 
$\pt$ spectra
in d+Au collisions 
 as published in Ref.~\cite{Adare:2013esx} for the 20\% most central collisions
via a simultaneous fit to all particle species.
We fit the spectra for $m_T - m<$~1~GeV/c and exclude $\pi^{\pm}$ below 0.5~GeV/c
because of possible larger contributions from resonance decays.
We fix $n$ = 1, corresponding to a linear boost profile.
The spectra overlaid with the best fit are shown in Figure~\ref{fig:spectra}.
The $\beta_{max}$ value is 0.70 and $T_{fo} =$139~MeV.  These fits describe
the data over the appropriate $p_T$ range 
to better than 10\%.
The statistical uncertainties on the points in the region of interest are small in comparison
to the systematic uncertainties.  To provide some estimate of the effect of 
these uncertainties on the extracted parameters we take the values of the uncertainties
which can change the shape of the particle spectra from Table IV in Ref.~\cite{Adare:2013esx}.
Since the correlation of the uncertainties as a function of $p_T$ is unknown we take them
to be maximally changing the slopes of the spectra in the range 0.5--3.0~GeV/$c$ by pivoting
the spectra around a point such that it is increased by the full systematic uncertainty at the
low (high) $p_T$ point and decreased by the full systematic uncertainty at the high (low) $p_T$
point.  This procedure provides the maximal variation in the blast-wave parameters within the 
measurement uncertainties and is thus likely to be an overestimate of the actual uncertainty.
Under this procedure we get $T_{fo}=$~145~MeV and $\beta_{max}=$~0.66 for the scenario where
the spectra are made steeper and $T_{fo}=$~127~MeV and $\beta_{max}=$~0.74 for the scenario
where the spectra are made flatter.  Blast-wave parameters for d+Au collisions
with $n$ not fixed to 1
are also reported in Ref.~\cite{Abelev:2008ab}.  

%($\beta_{max}$ and $T_{fo}$ values obtained here are
% slightly higher than those obtained 
%in Ref.~\cite{Abelev:2008ab} for the same collision system 
%and centrality.  However the PHENIX $p$ and $\bar{p}$ spectra have been
%corrected for feed-down from weak decays and the data in Ref.~\cite{Abelev:2008ab} have
%not.).

\begin{figure*}[ht]
\includegraphics[width=\textwidth]{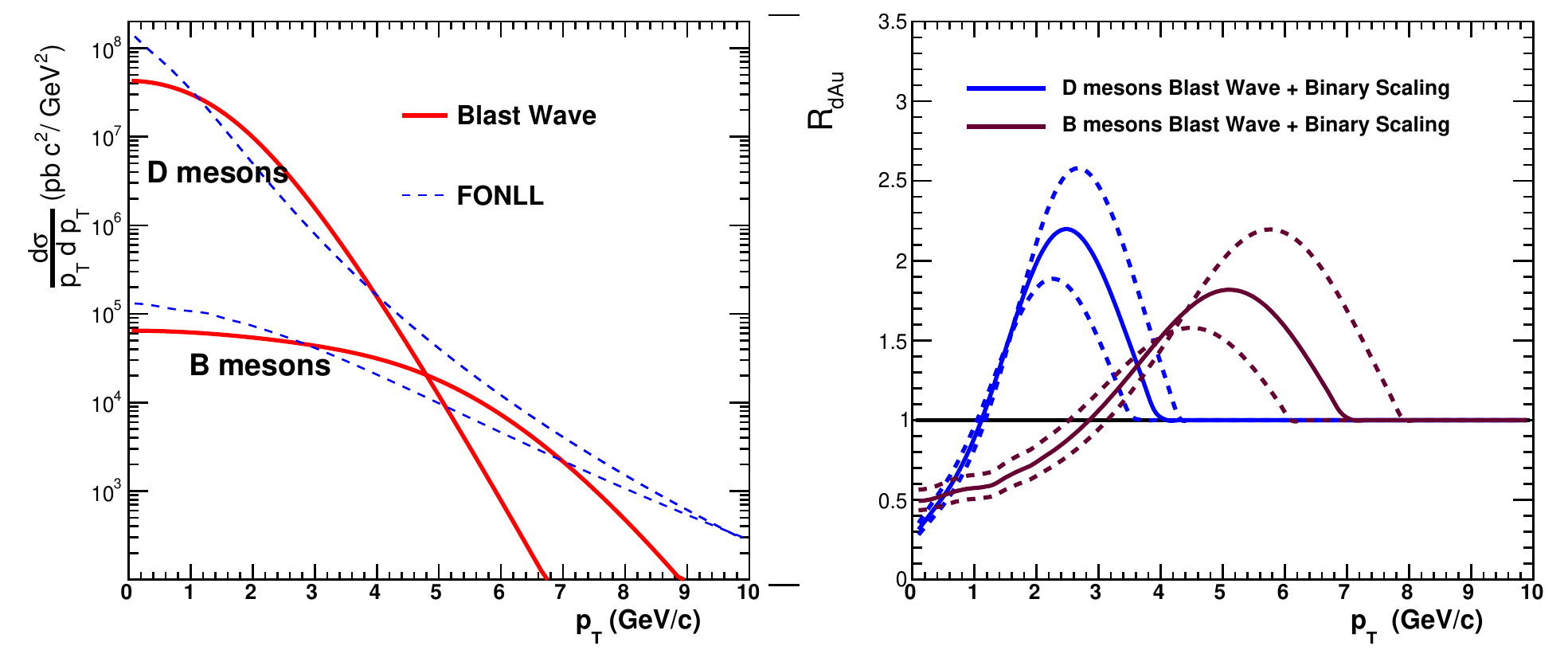}
\caption{(left)$D$ and $B$ meson $\pt$ spectra from FONLL~\cite{Cacciari:2001td,Cacciari:2012ny,fonllwp} 
(dashed lines)
and from the blast-wave calculation presented here (solid lines).  (right) The 
$R_{dA}$ from the comparison of the blast-wave and FONLL curves in the left panel with 
binary scaling added at high $\pt$.  The dashed lines show the changes in the blast-wave
expectations from the uncertainties on the blast-wave parameters discussed above.}
\label{fig:meson_RAA}
\end{figure*}

The blast-wave heavy meson spectra are determined from Eq.~\ref{eq:bw} using the 
parameters extracted above and the $D$ and $B$ meson masses (separately).
In order to quantify the enhancement for heavy mesons
expected from the blast-wave, we determine the $R_{dAu}$.  Here $R_{dAu}$ is the
ratio of the blast-wave heavy meson spectra divided by the expected heavy meson spectra from
the Fixed-Order-Next-to-Leading-Log (FONLL) calculation
of the heavy meson spectra in p+p
collisions~\cite{Cacciari:2001td,Cacciari:2012ny,fonllwp}.  
FONLL calculations compare well with a wide variety of heavy flavor data in 
p+p collisions at 
$\sqrt{s}=$~200~GeV~\cite{Adare:2009ic,Aggarwal:2010xp,Adare:2010de,Adamczyk:2012af}.
We use the default values from Ref.~\cite{fonllwp} of $m_c=$~1.5~GeV/$c^2$, $m_b=$~4.75~GeV/$c^2$
and $\mu_R=\mu_F = \sqrt{m^2 + p_T^2}$ where $\mu_R$ and $\mu_F$ are the renormalization
and factorization scales, respectively.
We normalize the blast-wave
spectra to have the same 
number of $D$ and $B$ mesons as the FONLL calculation.
The $\pt$ spectra for $D$ and $B$ mesons from FONLL and the blast-wave calculation
are shown in the left panel of Figure~\ref{fig:meson_RAA}.

We observed the blast-wave spectra to be below the FONLL spectra at
both low and high $\pt$.  They are greater than the FONLL spectra from
approximately 1--4~GeV/c for $D$ mesons and 2.5--7~GeV/c for $B$ mesons.
Regardless of the low $p_T$ physics, at high $p_T$ we
expect binary scaling of heavy mesons in d+Au collisions due to the dominance of hard physics
(unless other effects such as shadowing play a role).  
Mesons from a range
of momenta contribute to the electron spectrum at a given $p_T$. Therefore, in order
to have a sensible expectation for the electron $p_T$ spectra, we must include
mesons from a wide range of $p_T$ in the construction of the electron 
$R_{dAu}$~\cite{norm_footnote}.
At high $\pt$, in the calculations shown here when the blast-wave expectation 
decreases below the FONLL calculation,  we artificially enforce
binary scaling of the mesons (this is a 1\% change in the normalization for
$D$s and a 5\% change for $B$s).  The meson $R_{dAu}$, are shown for $D$ and $B$ mesons 
in the right panel of Figure~\ref{fig:meson_RAA}. We observe a large enhancement of $D$ mesons,
approximately a factor of two increase over FONLL at $\pt\approx$~2GeV/c.  We
observe a smaller enhancement of $B$ mesons, approximately a factor of 1.8 at around 5~GeV/c. 
The dashed lines in the figure show the variation of the expected meson $R_{dA}$
from the uncertainties on the blast-wave parameters extracted from the light hadron
data.  In all cases a large enhancement ($R_{dA}>$~1.5 is expected) for both
$D$ and $B$ mesons.  
When the blast-wave values from Ref.~\cite{Abelev:2008ab} are used, the
$D$ and $B$ meson $R_{dA}$ values have peaks at approximately the level
of the lower value of the uncertainties shown in Fig.~\ref{fig:meson_RAA} and
the peak positions are shifted to slightly higher $p_T$.
Since the blast-wave spectra are normalized to the integral
of $D$ and $B$ mesons from FONLL, the uncertainties on the FONLL calculation are
relevant only to the extent that shape of the spectra are changed.  
The uncertainties on the FONLL calculation are also available from Ref.~\cite{fonllwp}.
Comparisons with the data~\cite{Adare:2010de,Adamczyk:2012af} show that the data favors
the higher crosses sections within the FONLL uncertainty band.  
On the high
end of the systematic uncertainty band the FONLL calculation has an excess of low
$p_T$ particles compared to the central value.  This shape change increases the $R_{dA}$
of the mesons compared to the central values.

\begin{figure}
\includegraphics[width=\columnwidth]{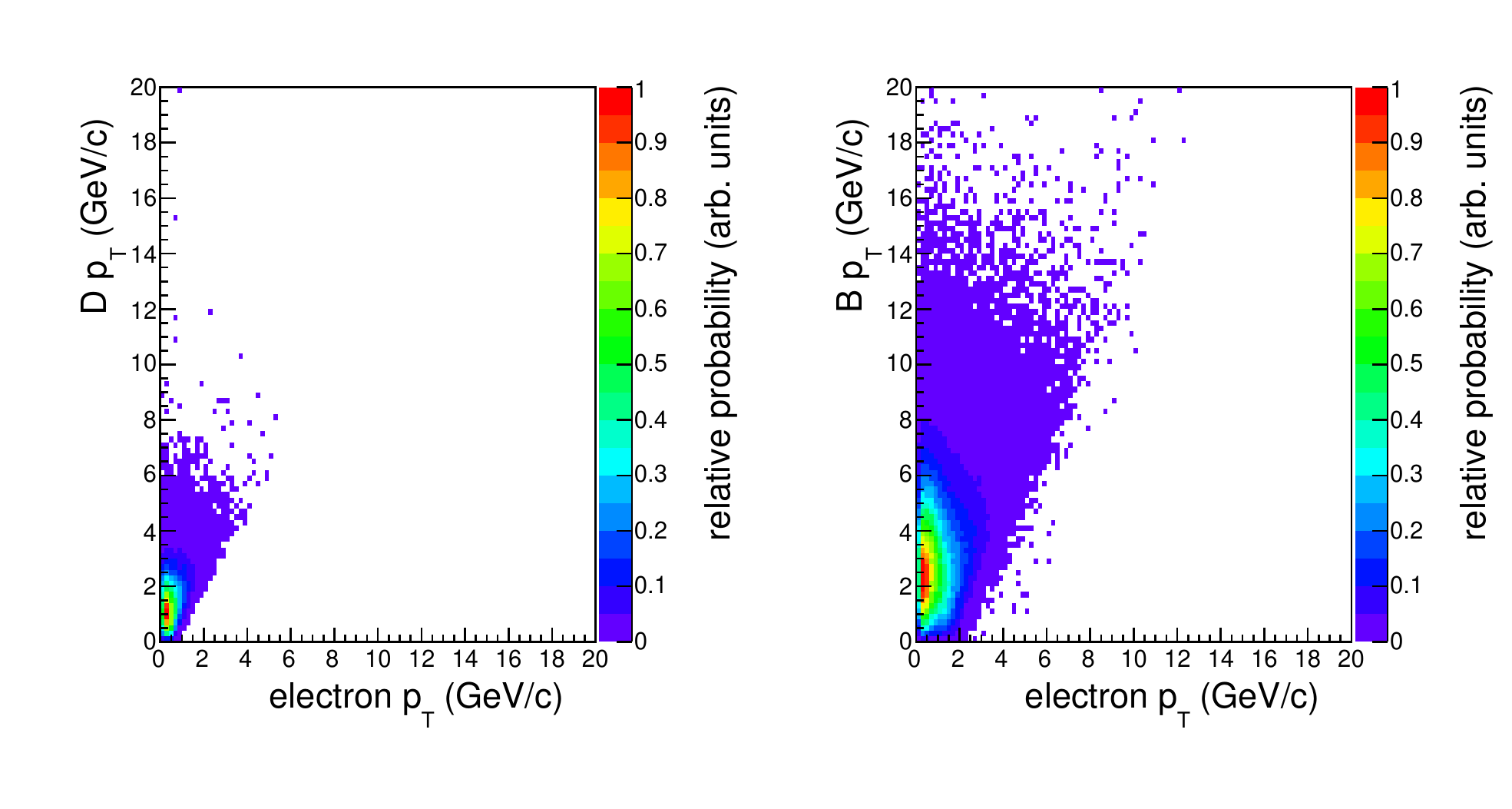}
\caption{The relative probability to for a heavy meson to decay into an electron at a given 
$p_T$. $D$ mesons are shown in the left panel and
$B$ mesons are shown in the right panel.  The decay kinematics are from PYTHIA8~\cite{Sjostrand:2007gs}.}
\label{fig:pt_corr}
\end{figure}

In order to determine the expected heavy flavor decay electron $R_{dAu}$ we use
PYTHIA (v 8.176)~\cite{Sjostrand:2007gs} to get the correlation between the $D$ or $B$ $\pt$ and the $p_T$
of the decay electron.
The correlations are shown in 
Figure~\ref{fig:pt_corr}.  The x-axis shows the $\pt$ of the electrons and
positrons (which are required to have $|\eta|<$~0.35 as in the experimental measurements)
and the y-axes have the $\pt$ of the parent $D$ and $B$ meson (decays in which a $B$ decays to a $D$
which subsequently decays to an electron are included in the $B$ meson plot).
We use the same procedure to extract the electron $\pt$ spectra for both the FONLL and blast-wave
meson $\pt$ spectra.  We take the branching ratios to be: $BR\left(B\to e\right)$ = 
10.86\%, $BR\left(B \to D \to e\right)$ 9.6\%
and $BR\left(D \to e\right)$ = 10.3\%~\cite{Beringer:1900zz}.

\begin{figure}
\includegraphics[width=\columnwidth]{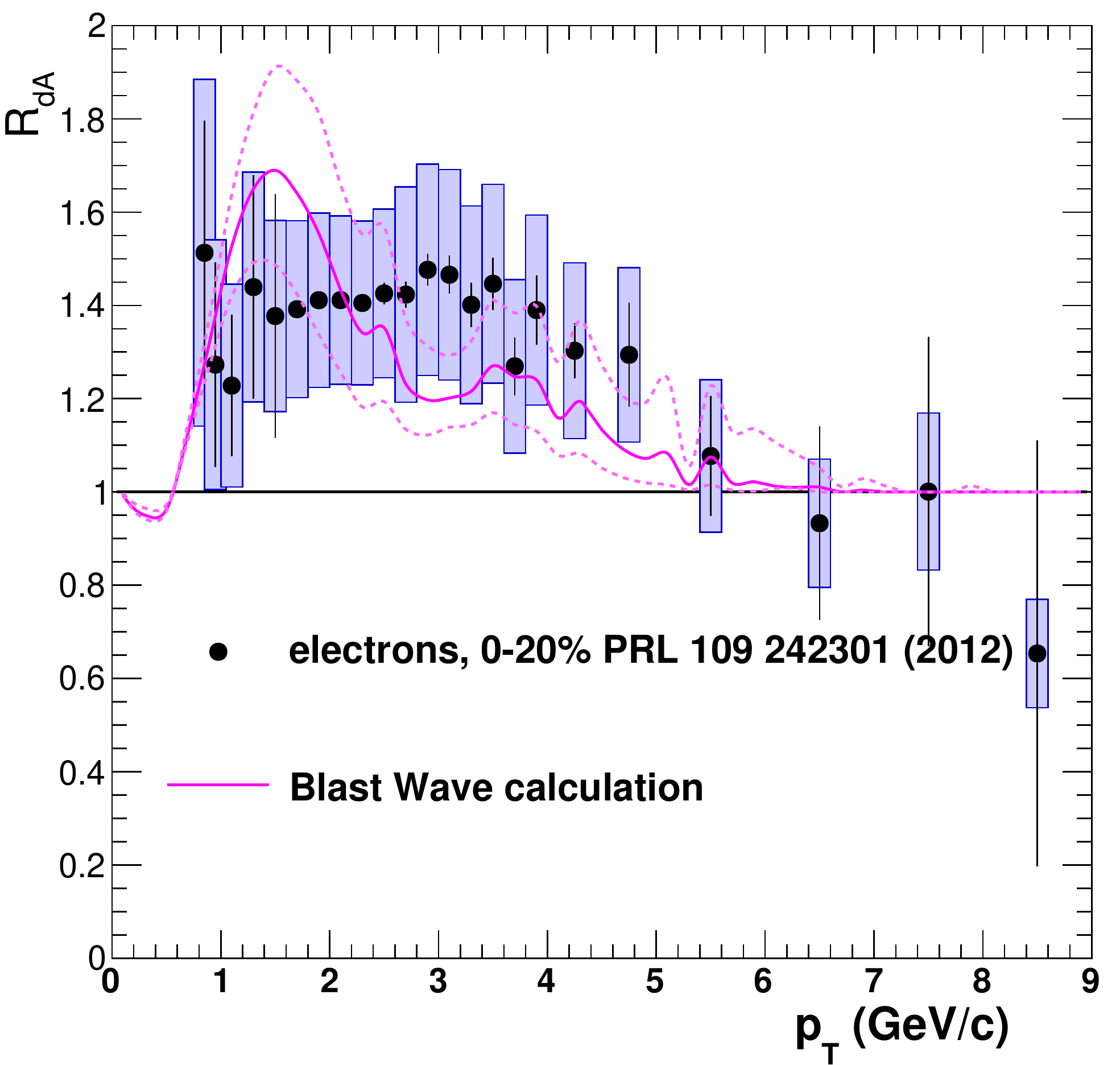}
\caption{The heavy flavor decay electron $R_{dA}$ for 0-20\% central d+Au collisions from 
Ref.~\cite{Adare:2012qb} (solid points) and from the blast-wave calculations presented in this work
(curve).  The dashed lines show the changes in the blast-wave expectations from the
uncertainties on the blast-wave parameters discussed above.}
\label{fig:results}
\end{figure}

The results for the electron $R_{dAu}$  are shown in Figure~\ref{fig:results}
overlaid with the measured electron $R_{dAu}$~\cite{Adare:2012qb}.  
The uncertainties on the data are large
and the uncertainties on the blast-wave calculation are shown as the dashed lines.
The magnitude
of the enhancement expected from the blast-wave calculation is in good agreement
with the data.  At $\pt\approx$~1-2.5~GeV/c there is a peak in the calculation 
that is not seen in the data.  At high $\pt$, both the data and the calculation, by construction,
approach unity at high $p_T$ in a similar manner.  
 However,
since the blast-wave calculation qualitatively reproduces the data, if a hydrodynamic
description of the light hadrons is valid in d+Au collisions, then it is possible that
the same is also true for heavy flavor.

\section{Predictions for p+Pb Collisions at 5.02~TeV}

It is, of course, natural to ask whether this effect could also play a role in p+Pb collisions
at the LHC.  The ALICE collaboration has published the results of blast-wave fits to 
identified particle spectra in multiplicity classes in Ref.~\cite{Abelev:2013haa}.  
The blast-wave implementation that they have used is slightly different from the one used above and 
allows $n$ to be a fit parameter as well.
Preliminary results on the $R_{pPb}$ exist from ALICE for both reconstructed $D$ mesons
and electrons from the decay of heavy mesons~\cite{Heide:2013qsa,fortheALICE:2013ica}.  
In both cases the $R_{pPb}$ is consistent with unity;  however the uncertainties
are large.  The $D$ meson measurement is compatible with a 10-20\% enhancement.
The heavy flavor decay electron measurement has a central value of approximately 30\% enhancement
for $p_T<6$~GeV/c with uncertainties of approximately the same size.

\begin{figure}
\centering
\includegraphics[width=\columnwidth]{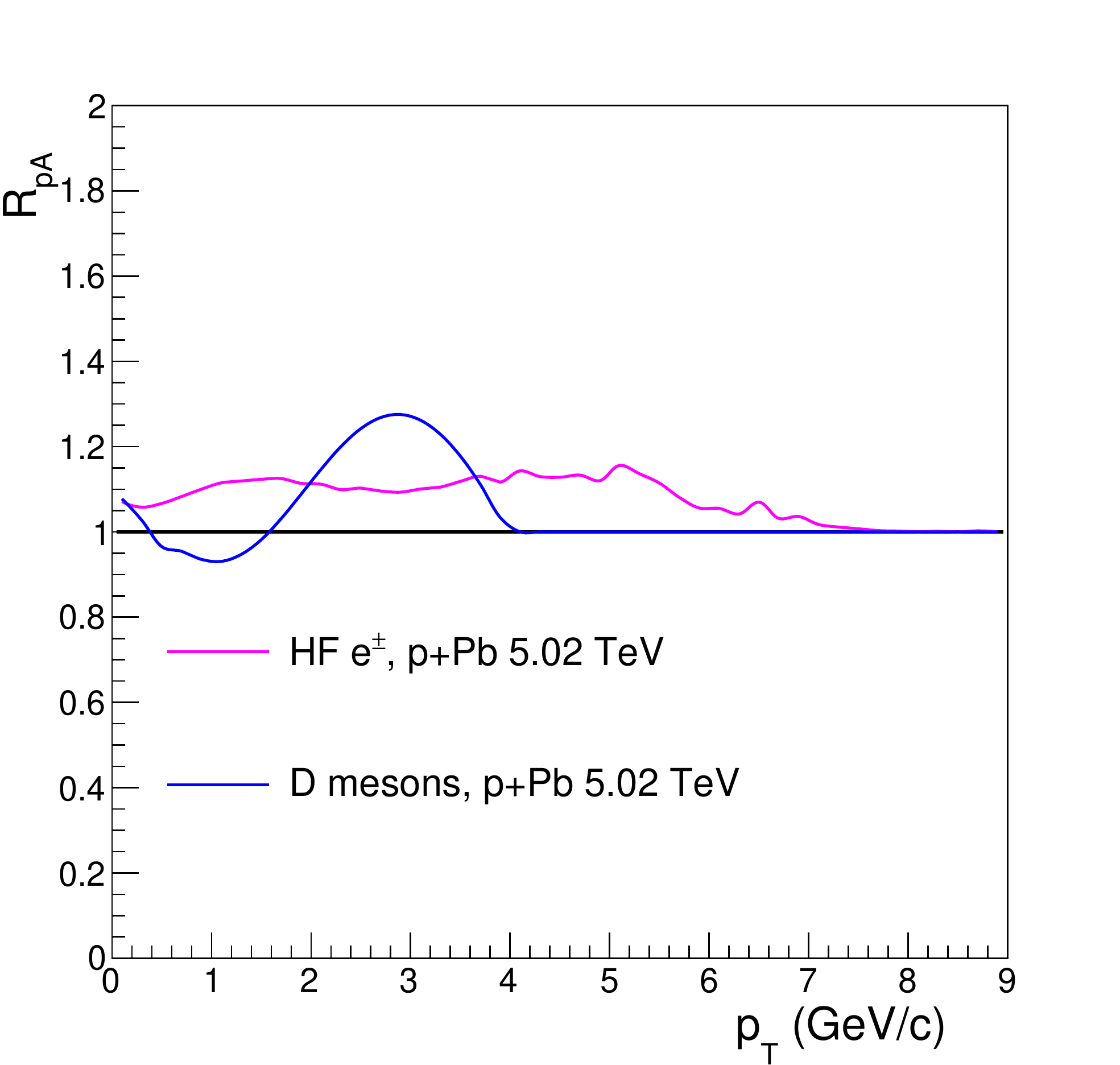}
\caption{Predictions for p+Pb collisions at 5.02~TeV.  Blast-wave fit results
from the 5\% highest multiplicity 
 p+Pb collisions~\cite{Abelev:2013haa} and FONLL~\cite{Cacciari:2012ny,fonllwp} heavy meson
spectra have been used to generate these
results.}
\label{fig:lhc}
\end{figure}

Unfortunately,
the heavy flavor results currently only are measured
 for minimum bias data so a direct comparison of the predictions
with data is not possible.  As an illustration of the possible magnitude of the effect, we
have used the blast-wave parameters from the highest multiplicity bin, 0-5\%, and FONLL
calculations for 5.02~TeV~\cite{Cacciari:2012ny,fonllwp}.  Results for electrons and $D$ mesons
are shown in Figure~\ref{fig:lhc}.  The $D$ meson enhancement reaches a maximum
of approximately 20\% at $\pt\approx$~3~GeV/c and the electrons are enhanced by 
10--20\% nearly independently of $\pt$ over the range of 1--6~GeV/c.  The calculations are
for the highest multiplicity event class and show larger modifications than what would be expected for
minimum bias collisions.
Because of the harder $D$ and $B$ meson
$p_T$ spectra at the higher collision energy there is a smaller enhancement
of heavy flavor mesons than at RHIC, despite the larger maximal velocity extracted
from the blast-wave fits.

\section{Conclusions}

Given the large uncertainties on the available heavy flavor data in d+Au collisions at RHIC
and the large uncertainties on the blast-wave calculation here, it is important to consider
how a radial flow interpretation of heavy flavor data in very small collision systems would be
verified or ruled out.  The clearest evidence will come from charm and bottom separated results
being made possible by recently installed vertex detectors at both STAR and PHENIX.
Reconstructed $D$ mesons from STAR and charm and bottom separated electron measurements 
from PHENIX will show the meson mass dependence of the heavy flavor enhancement seen by
PHENIX~\cite{Adare:2012qb}.  Additionally, it is of interest to study multiplicity selected
p+p collisions at $\sqrt{s}$~=~200~GeV in order to investigate how the blast-wave parameters
evolve with both collision system and event activity and how that informs the interpretation
of the d+Au data discussed here in terms of radial flow.

Recently, there has been much interest in the possibility of hydrodynamic flow in very small
collisions systems.  Here we have raised the possibility that the enhancement of 
heavy flavor decay electrons previously observed~\cite{Adare:2012qb}
could be caused by radial flow using a blast-wave parameterization constrained by the light hadron
data.  We find qualitative agreement between the data and the prediction of this
model, suggesting hydrodynamics as one possible explanation of the enhancement
of electrons from heavy flavor decay observed in d+Au collisions.
Further measurements have the potential to constrain any possible role of hydrodynamics in very small collision
systems.  $D$ meson spectra at RHIC are especially interesting as the modifications should be
significantly larger than is seen at the LHC.

\section{Acknowledgments}
The author thanks Dave Morrison, Paul Stankus and Jamie Nagle for helpful conversations.
The author is supported by the U.S. Department of Energy under
contract DE-AC02-98CH10886. 
\bibliography{spectra_dAu}

\end{document}